# Quantum gravity and spin systems[*]

W. Beirl, P. Homolka[†], B. Krishnan, H. Markum and J. Riedler

Institut für Kernphysik, Technische Universität Wien, A-1040 Vienna, Austria

A new method for nonperturbative investigations of quantum gravity is presented in which the simplicial path integral is approximated by the partition function of a spin system. This facilitates analytical and numerical computations considerably. In two dimensions equivalence to an Ising model with ternary couplings is recovered. First simulations in four dimensions indicate strong similarities to the phase structure of original Regge theory.

More than thirty years ago Regge developed a discrete description of General Relativity in which space-time is triangulated by a simplicial lattice, the Regge skeleton [1]. Thus, the lattice itself becomes a dynamical object, with the quadratic edge lengths $q$ describing the evolution of space-time. Within this scheme the Einstein-Hilbert action for lattice gravity with cosmological constant $\lambda$ in $d$ dimensions is given by

$$I(q) = \lambda \sum_{s^d} V(s^d) - 2\beta \sum_{s^{d-2}} \delta(s^{d-2}) V(s^{d-2}), \quad (1)$$

where the first sum is over all $d$-simplices $s^d$ in the simplicial complex and $V(s^d)$ is the $d$-volume of the indicated simplex. The curvature of the lattice is concentrated on the $(d-2)$-simplices leading to deficit angles $\delta(s^{d-2})$. $\beta$ denotes the gravitational coupling.

In the path integral formulation a quantization of the above action proceeds by evaluating the expression

$$Z = \int D[q] e^{-I(q)} . \quad (2)$$

Unfortunately, a unique prescription for the measure does not exist, however, an appropriate choice proposed in the literature is [2]

$$\int D[q] = \prod_l \int \frac{dq_l}{q_l^m} \mathcal{F}(\{q_l\}), \quad (3)$$

[*]Supported in part by "Fonds zur Förderung der wissenschaftlichen Forschung" under Contract P9522-PHY.
[†]Present address: Dépt. de Physique, Université Laval, Québec G1K 7P4, Canada.

with $m \in \mathbb{R}$ defining a one-parameter family and $\mathcal{F} = 1$ for Euclidean configurations of squared edge lengths $\{q_l\}$ but $\mathcal{F} = 0$ otherwise.

The central idea of this investigation is to transform the quantum gravity path integral to a partition function of a spin system [3,4]. The model is defined by allowing the squared edge lengths to take on only two values

$$q_l = 1 + \epsilon \sigma_l , \quad 0 \leq \epsilon < \epsilon_{max} , \quad \sigma_l \in Z_2 , \quad (4)$$

similar to the Regge-Ponzano model [5]. The real parameter $\epsilon$ is restricted to fulfill the Euclidean triangle inequalities for the $q_l$'s so that all $2^{N_1}$ configurations are allowed ($N_1$ is the total number of edges), i. e. $\mathcal{F} = 1$ for all configurations $\{q_l\}$.

Using (4) the measure (3) can be replaced by

$$\sum_{\sigma_l = \pm 1} \exp[-m \sum_l \ln(1 + \epsilon \sigma_l)] =$$
$$= \sum_{\sigma_l = \pm 1} \exp[-N_1 m_0(\epsilon) - \sum_l m_1(\epsilon) \sigma_l], \quad (5)$$

with $m_0 = -\frac{1}{2} m \epsilon^2 + O(\epsilon^4)$ and $m_1 = m(\epsilon + \frac{1}{3} \epsilon^3) + O(\epsilon^5)$.

In two dimensions, according to the Gauss-Bonnet theorem, the Einstein-Hilbert action as well as its discretized version is a topological invariant. Therefore, it can be dropped if one considers manifolds with fixed topology. The only remnant of the gravitational action (1) is the cosmological term. In the following the manifolds will be restricted to toroidal topology. The building blocks of a 2-dimensional simplicial lattice are triangles and thus $V(s^d) = V(s^2) = A_t$. To rewrite the action $I = \lambda \sum_t A_t$ in terms of $\sigma_l$ we consider a single triangle $t$ with squared edge lengths $q_1, q_2, q_l$. Its area can be expressed as

$$A_t = \left| \begin{matrix} q_1 & \frac{1}{2}(q_1 + q_2 - q_l) \\ \frac{1}{2}(q_1 + q_2 - q_l) & q_2 \end{matrix} \right|^{\frac{1}{2}} =$$
$$= [\frac{3}{4} + \frac{1}{2}(\sigma_1 + \sigma_2 + \sigma_l)\epsilon +$$
$$+ \frac{1}{2}(\sigma_1\sigma_2 + \sigma_1\sigma_l + \sigma_2\sigma_l - \frac{3}{2})\epsilon^2]^{\frac{1}{2}} . \quad (6)$$

Expanding $A_t$ the series consists only of terms up to $\sigma^3$ since $\sigma_i^2 = 1$. This suggests the following ansatz

$$A_t = c_0(\epsilon) + c_1(\epsilon)(\sigma_1 + \sigma_2 + \sigma_l) + c_2(\epsilon)(\sigma_1\sigma_2 +$$
$$+ \sigma_1\sigma_l + \sigma_2\sigma_l) + c_3(\epsilon)\sigma_1\sigma_2\sigma_l . \quad (7)$$

There are only four possible values for the area of a triangle. Computing these areas and comparing with (7) results in exact solutions for the coefficients $c_i$, e.g.

$$c_2 = \frac{1}{16}[2\sqrt{3} - \sqrt{(1-\epsilon)(3+5\epsilon)} -$$
$$- \sqrt{(1+\epsilon)(3-5\epsilon)} \, ] . \quad (8)$$

Obviously one must have $\epsilon < \frac{3}{5} = \epsilon_{max}$ for the triangle areas to be real and positive. Inserting (5) and (7) into the partition function yields

$$Z = \sum_{\sigma_l = \pm 1} J \exp\{-\sum_l (2\lambda c_1 + m_1)\sigma_l -$$
$$- \lambda \sum_t [c_2(\sigma_1\sigma_2 + \sigma_1\sigma_l + \sigma_2\sigma_l) +$$
$$+ c_3\sigma_1\sigma_2\sigma_l]\}, \quad (9)$$

with $J = \exp(-\lambda N_2 c_0 - N_1 m_0)$ and $N_2$ the total number of triangles. Thus, the path integral becomes the partition function of a system consisting of a spin $\sigma_l$ at each edge $l$, with an external "magnetic field" and with 2- and 3-spin nearest neighbor interactions. Assigning the spin to the corresponding edge of the originally triangular lattice and drawing the interactions as lines, a Kagomé lattice is obtained.

The term linear in $\sigma$ can be removed by a convenient choice of the measure, $m_1 = -2\lambda c_1$, and for $\epsilon \to 0$ the 3-spin couplings $c_3$ can be neglected compared to $c_2$. Hence, the remaining partition function is that of an Ising model on a Kagomé lattice

$$Z_I = \sum_{\sigma_l = \pm 1} J \exp\{Q \sum_l (\sigma_1 + \sigma_2 + \sigma_3 + \sigma_4)\sigma_l\}, \quad (10)$$

where $\sigma_{1,2}$ and $\sigma_{3,4}$ correspond to edges of neighboring triangles having $\sigma_l$ in common. To compute the critical coupling $Q_c = (-\frac{1}{2}c_2\lambda)_c$ one uses the fact that $Z_I$ is connected to the partition function of a decorated honeycomb lattice via a triangle-star transformation and further to a conventional honeycomb lattice by a decoration-iteration transformation [6]. After all, one gets

$$\lambda_c = -2\frac{0.4643}{c_2} , \quad (11)$$

which means that the antiferromagnetic regime, $\lambda > 0$, is governed by frustration, whereas for ferromagnetic interaction, $\lambda < 0$, the system undergoes a $2^{nd}$-order phase transition. Thus, close to the critical point our model with $c_3 = 0$ and $m_1 = -2\lambda c_1$ is equivalent to a $\varphi^4$ field theory. On the other hand, 2d-gravity in the continuum is known to be equivalent to the Liouville theory. An analysis of our model (9) for reasonable measure and couplings is necessary to shed light on the connection with continuum theory.

A generalization of the 2-dimensional action to higher dimensions becomes rather complicated. In two dimensions only the three edges of a triangle are coupled in the action, whereas in three dimensions one is faced with terms of $6^{th}$ order at least, and with additional contributions from the Regge action. Recently, the Ising-link model in three dimensions has been analyzed via mean field techniques as well as with Monte Carlo methods [4]. However, critical exponents and the behavior of the curvature at the transition point $\beta_c$ seem to differ from what was found for the Regge theory with continuously varying edge lengths.

The situation is even more complicated in four dimensions where one has to deal with 10 edges in each simplex. Nevertheless, numerical simulations can be made very efficient using the heat bath algorithm. In the actual computations we replaced (4) by $q_l = b_l(1 + \epsilon\sigma_l)$ allowing the lattice to fluctuate around flat space. A 2-dimensional simplicial lattice with equilateral links can be embedded in flat space, but for the 4-dimensional case $b_l$ has to take different values depending on the type of the edge $l$. In particular $b_l = 1, 2, 3, 4$ for edges, face diagonals, body diagonals, and the hyperbody diagonal of a hypercube, respectively. Preliminary studies show an interesting phase structure.

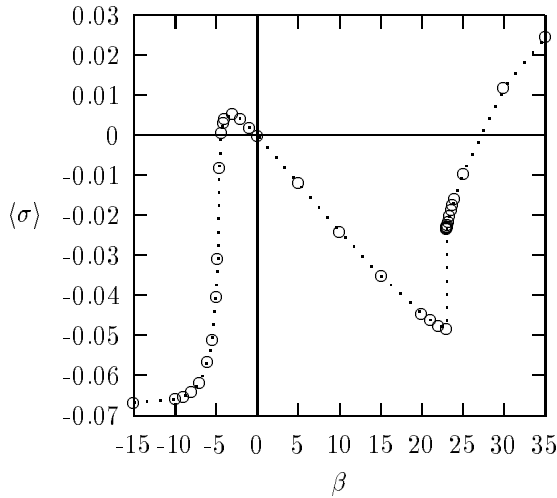

Figure 1. Spin expectation value as a function of $\beta$ from simulations of spin quantum gravity on a toroidal $8^4$ lattice.

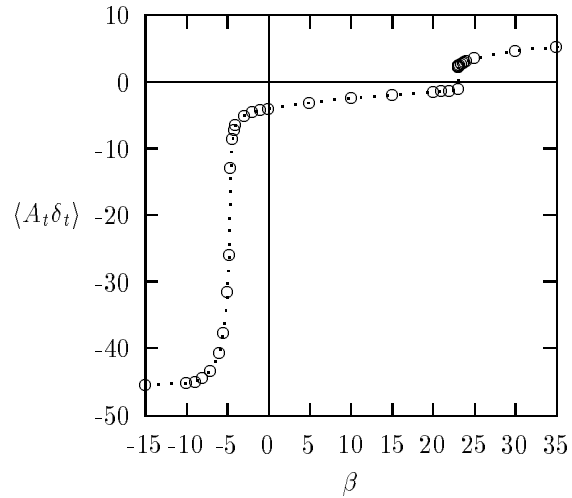

Figure 2. Average Regge action for different couplings $\beta$. Uniform measure $m = 0$, cosmological constant $\lambda = 0$ and $\epsilon = 0.0875$ are used.

The spin expectation value of a 4-dimensional toroidal lattice with $8^4$ vertices is depicted in Fig. 1. The parameters were set to $\lambda = m = 0$, $\epsilon = 0.0875$ and the range of $\beta$ is extended to negative values. The system undergoes phase transitions at two different points, first in the negative region at $\beta \approx -4.67$ and second at a large positive coupling $\beta \approx 23.01$.

In Fig. 2 the expectation value of the corresponding Regge action as a function of $\beta$ is displayed. The resemblance of the curve for $\beta > 0$ to that obtained from continuously varying edge lengths is very encouraging, cf. [2]. One has to determine the value of the critical points and the order of the phase transitions in the limit of infinitely many vertices, $N_0 \to \infty$, in detail. Recent simulations indicate that both, the spin gravity model and Regge theory, belong to the same universality class. If this is indeed the case one could perform calculations much easier than with conventional Regge calculus.